# Decay rate enhancement of diamond NV-centers on diamond thin films

Hao Li,* Jun-Yu Ou, Vassili A. Fedotov and Nikitas Papasimakis

*Optoelectronics Research Center & Center for Photonic Metamaterials, University of Southampton, Southampton SO17 1BJ, UK*

*Correspondence to: H.Li@soton.ac.uk

We demonstrate experimentally two-fold enhancement of the decay rate of $NV^0$ centers on diamond/Si substrate as opposed to a bare Si substrate. We link the decay enhancement to the interplay between the excitation of substrate modes and the presence of non-radiative decay channels. We show that the radiative decay rate can vary by up to 90% depending on the thickness of the diamond film.

The nitrogen-vacancy (NV) defect in diamond constitutes an important test ground and building block for quantum devices[1-7]. Structured plasmonic films, particles, and hyperbolic metamaterials are known to enhance decay rates of quantum emitters owing to the Purcell effect[8-14]. Charge transfer from silicon and metallic substrates provides an additional mechanism to control NV center emission[15], while the sensitivity of NV centers to the local dielectric environment has been used for mapping of local density of states[16] and near-field microscopy[17]. However, the NV center decay rates are broadly distributed, which hinders studies of their interactions with their environment[18, 19].

In this paper, we report a two-fold increase of the decay rate for ensembles of $NV^0$ emitters resting on a thin diamond film, as compared to a bare silicon substrate, which we observed experimentally using time-resolved cathodoluminescence (TR-CL). We attribute such an increase to the existence of non-radiative channels and coupling of the emitters to Fabry-Perot modes of the thin film.

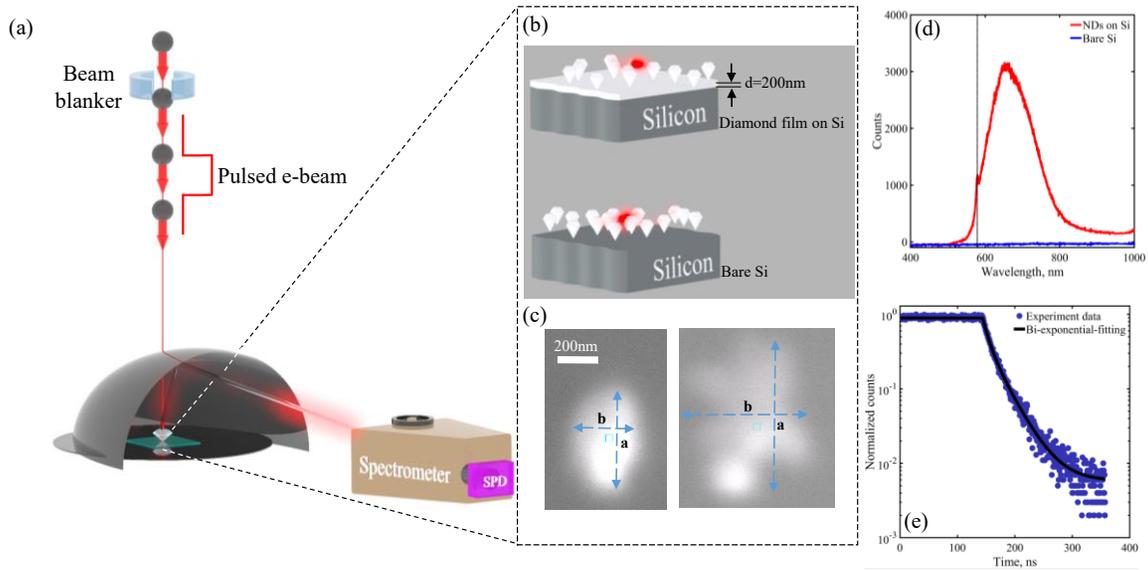

**Fig. 1.** (a) Schematic of the scanning electron microscope-based system for time-resolved electron-induced light emission spectroscopy. Electrons impinge on the sample through a small hole in a parabolic mirror, which collects and collimates the emitted light. The beam is subsequently directed to the spectrometer. At the output of the spectrometer, a single photon detector (SPD) is used to detect the emitted photons. (b) The samples consist of NDs deposited on two different substrates: i) a 200 nm thick diamond film on 500 μm thick silicon and ii) a 500 μm thick silicon substrate. (c) SEM images of ND clusters are shown at the bottom. The blue rectangle marks the position of the electron beam during TR-CL measurements. The dimensions of the cluster are indicated by the longer and shorter edge lengths, *a* and *b*, respectively. (d) CL spectrum of NVs on Si (red line) and background signal (blue line). The dashed line indicates the wavelength zero phonon line. (e) TR-CL trace of NVs on Si (blue circles) and bi-exponential fitting (black line) (see Supporting Information Section S2).

In our experiments, we considered ~120 nm large nanodiamonds containing ~$10^3$ NV centers. The nanodiamonds were diluted in methanol and the solution was mixed for 10 min in an ultrasonic bath. The nanodiamonds were then deposited by drop casting on two different substrates: a 200 nm thick diamond film on silicon and a bare silicon wafer (see Fig. 1(b)). We characterized NV centers in the deposited nanodiamonds at room temperature using a scanning electron microscope (SEM) operating in fixed-spot mode. The electron beam was incident on the samples through a small hole in a parabolic mirror, which collected and collimated the emitted light. The collected light was subsequently directed to the entrance of a VIS/NIR spectrometer, as shown in Fig. 1 (a). The spectrometer selected photons within a 3.6 nm wavelength range around the central wavelength of 575 nm and directed them to the input of a single photon detector. The beam blanker driven with a wave function generator produced a pulsed electron beam and also provided the synchronization required to implement time-correlated single-photon counting[20]. A beam current of ~1.8 nA was maintained in each TR-CL measurement. Our system can measure reliably lifetimes as short as ~3 ns.

We selected 60 nanodiamond clusters on each type of substrate with sizes in the range 120-1500 nm (see Fig. 1(c) and Supporting Information S3) and studied the emission of the neutral $NV^0$ centers at the zero phonon line, as shown in Fig. 1(d). Thus, only photons in the range of 575 $\pm$ 1.8 nm were selected at the input of the single photon detector. Figures 2(a-b) present the distribution of the measured lifetimes of NV centers deposited on a bare Si substrate and on a diamond film, respectively. We characterized the statistical distribution of the lifetimes by the corresponding average ($\mu$) and the standard deviation ($\sigma$). NV centers on bare Si substrate are seen to exhibit long lifetimes ($\mu$=30 ns) and a broad distribution ($\sigma$=6 ns). On the other hand, for NV centers on the diamond film, we observe a substantial shortening of the average lifetime ($\mu$=17 ns) accompanied by a narrower distribution ($\sigma$=4 ns).

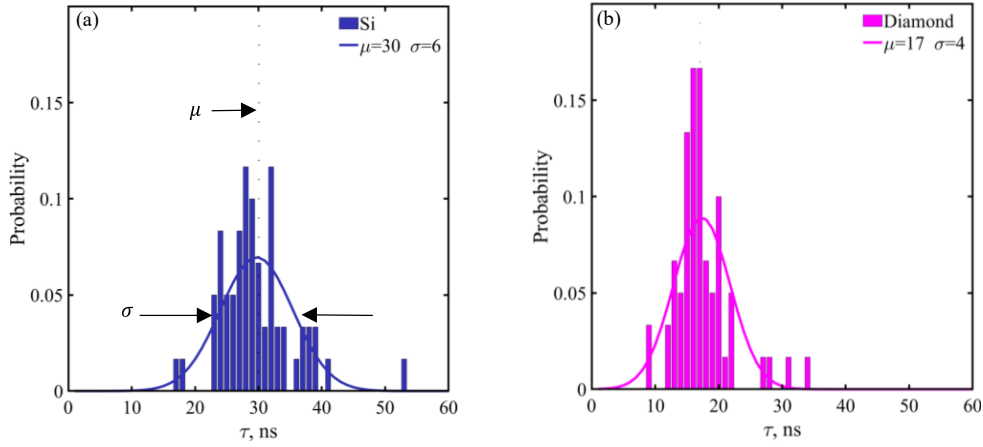

**Fig. 2.** Lifetime distributions of NV$^0$ centers in diamond nanoparticles deposited on silicon (a) and on diamond thin film on silicon (b). The histograms are obtained from measurements of 60 different ND clusters on each substrate and are fitted by a Gaussian distribution, where μ is the mean (dotted vertical lines) and σ is the standard deviation (in ns).

We argue that the changes in the lifetime distribution observed for NV centers on a diamond film are a result of strong interactions between diamond nanoparticles and guided optical modes in the diamond film. To demonstrate this we performed full-wave 3D electromagnetic modeling of the emission from a dipole embedded in a diamond nanoparticle placed on a diamond film of varying thickness. The modeling also allowed us to distinguish between radiative and non-radiative decay channels. A comparison of computationally obtained and experimentally measured total lifetimes is presented in Table 1. In agreement with our experimental results, our simulations show that the lifetime is shorter in the presence of a thin diamond film (81 ns) than in the case of a bare Si substrate (91 ns). We attribute the difference between the values of experimental and numerical lifetimes mainly to the fact that in our numerical calculations we considered single nanodiamond particles, whereas experimental measurements involved clusters of nanodiamonds.

In particular, our numerical results indicate that the faster decay rates obtained for thin diamond films are due mainly to non-radiative loss in the underlying bulk Si (see Supporting Information table S1). Importantly, the enhancement or suppression of radiative and non-radiative decay rate depends strongly on the film thickness (see Fig. 3(a)), which indicates coupling to slab modes in the film[21]. Indeed, at the thickness of the experimentally measured samples (90 nm) radiative (non-radiative) decay rates are suppressed (enhanced), whereas at 160 nm film thickness the situation is reversed. This is further illustrated in radiated field maps plotted in Figs. 3(b) and 3(c) for the film thickness of 90 nm and 160 nm, respectively. The modelled dependence of the radiative decay rate on the film thickness can be fitted with a sine function: $k + l\sin(\omega x + m)$, as shown in Fig. 3 (a). From the fitting parameters we obtain the period of the sine function, $T_{\text{diamond}}^{\text{rad}} = \frac{2\pi}{\omega_{\text{diamond}}^{\text{rad}}}$ =126 nm, which is close to the thickness of a diamond film supporting the fundamental Fabry-Perot mode at λ=575nm: $T_{\text{diamond}}^{\text{FP}} = \frac{575 \text{ nm}}{2n_{\text{diamond}}} = 120$ nm. We, therefore, conclude that the decay rate can be efficiently controlled by changing the thickness of the supporting thin film. In particular, the tuning depth of NV radiative decay rate for a thin diamond film can be as high as $\frac{2l_{\text{diamond}}^{\text{rad}}}{k_{\text{diamond}}^{\text{rad}}} =$ 89%, where $l_{\text{diamond}}^{\text{rad}}$ and $k_{\text{diamond}}^{\text{rad}}$ are the fitting parameters of the sine function. Finally, we argue that in our study the average lifetime is largely independent of the inhomogeneities in the substrate and immediate dielectric environment, as well as NV emitter position and orientation of its electric dipole moment. Indeed, the large number of NV emitters per particle yields a Purcell factor effectively averaged over different positions and dipole orientations of the emitters in nanodiamonds. Furthermore, our measurements involved clusters of different sizes and aspect ratios with no strong correlation observed between the cluster size/shape and decay rate (see

Supporting Information section S3).

**Table 1.** Experimentally measured ($\bar{\tau}_{Exp}$) and numerically simulated ($\tau_{Sim}$) NV center total lifetimes on diamond film and bare silicon substrate. Experimental values correspond to the average of the distribution.

| Substrates | $\bar{\tau}_{Exp}(ns)$ | $\tau_{Sim}(ns)$ |
|---|---|---|
| Si | 30 | 91.49 |
| Diamond | 17 | 81.73 |

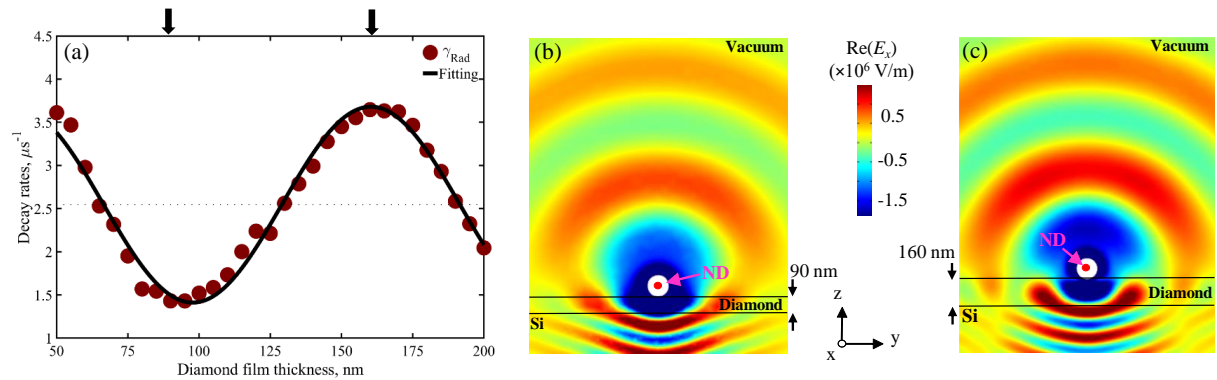

**Fig. 3.** (a) Modelled radiative decay rates of an NV center on a diamond film as a function of the film thickness. The electric dipole moment of the emitter is oriented parallel to the plane of the substrate. The solid line corresponds to a sine function fit to the numerically calculated decay rates (solid circles), while the dashed line corresponds to the radiative decay rate of an NV center on a bare silicon substrate. (b-c) The real part of $x$-component of the electric field emitted in $zy$-plane by an NV center of a nanodiamond (ND) sitting on a diamond film with a thickness of 90 nm and 160 nm, respectively (the electric dipole moment of the NV center is oriented along $x$-axis).

In conclusion, we experimentally demonstrate two-fold enhancement of the decay rate of $NV^0$ centers in nanodiamonds when deposited on a thin diamond film. We attribute the enhancement to the presence of non-radiative decay channels and coupling of $NV^0$ centers to the optical slab modes supported by the film, which leads to increase of both radiative and non-radiative decay rates. As such, we show that varying the thickness of the diamond film allows tuning of the radiative decay rate by up to 90%. Our study provides insights into the mechanism of Purcell enhancement of NV emission from nanodiamonds deposited on thin films and puts forward simple means of controlling the corresponding emission statistics.

The authors would like to acknowledge financial support from the Engineering and Physical Sciences Research Council UK (Grants No. EP/N00762X/1 and No. EP/M0091221), China Scholarship Council (No. 201608440362). Following a period of embargo, the data from this paper will be available from the University of Southampton ePrints research repository: http://doi.org/10.5258/SOTON/DXXX.

## References

1. I. Aharonovich and E. Neu, Advanced Optical Materials **2** (10), 911-928 (2014).
2. A. B. R. Brouri, J.-. Poizat, and P. Grangier, Optics Letters **25** (17) (2000).
3. S. M. C. Kurtsiefer, P. Zarda, and H. Weinfurter, Phys Rev Lett **85** (2) (2000).
4. K. M. C. Fu, C. Santori, P. E. Barclay, I. Aharonovich, S. Prawer, N. Meyer, A. M. Holm and R. G. Beausoleil, Applied Physics Letters **93** (23) (2008).
5. A. Huck, S. Kumar, A. Shakoor and U. L. Andersen, Phys Rev Lett **106** (9), 096801 (2011).
6. J. T. Choy, I. Bulu, B. J. M. Hausmann, E. Janitz, I. C. Huang and M. Lončar, Applied Physics Letters **103** (16) (2013).
7. H. Siampour, S. Kumar and S. I. Bozhevolnyi, ACS Photonics **4** (8), 1879-1884 (2017).
8. M. B. S. Schietinger, T. Aichele, and O. Benson, Nano Lett **9** (4), 1694-1698 (2009).
9. J. T. Choy, B. J. M. Hausmann, T. M. Babinec, I. Bulu, M. Khan, P. Maletinsky, A. Yacoby and M. Lončar, Nature Photonics **5** (12), 738-743 (2011).
10. A. S. Zalogina, R. S. Savelev, E. V. Ushakova, G. P. Zograf, F. E. Komissarenko, V. A. Milichko, S. V. Makarov, D. A. Zuev and I. V. Shadrivov, Nanoscale **10** (18), 8721-8727 (2018).
11. S. Kumar, A. Huck and U. L. Andersen, Nano Lett **13** (3), 1221-1225 (2013).
12. S. K. H. Andersen, S. Kumar and S. I. Bozhevolnyi, Nano Lett **17** (6), 3889-3895 (2017).
13. S. I. Bogdanov, M. Y. Shalaginov, A. S. Lagutchev, C. C. Chiang, D. Shah, A. S. Baburin, I. A. Ryzhikov, I. A. Rodionov, A. V. Kildishev, A. Boltasseva and V. M. Shalaev, Nano Lett **18** (8), 4837-4844 (2018).
14. M. Y. Shalaginov, S. Ishii, J. Liu, J. Liu, J. Irudayaraj, A. Lagutchev, A. V. Kildishev and V. M. Shalaev, Applied Physics Letters **102** (17) (2013).
15. S. Stehlik, L. Ondic, A. M. Berhane, I. Aharonovich, H. A. Girard, J.-C. Arnault and B. Rezek, Diamond and Related Materials **63**, 91-96 (2016).
16. A. W. Schell, P. Engel, J. F. Werra, C. Wolff, K. Busch and O. Benson, Nano Lett **14** (5), 2623-2627 (2014).
17. J. Tisler, T. Oeckinghaus, R. J. Stohr, R. Kolesov, R. Reuter, F. Reinhard and J. Wrachtrup, Nano Lett **13** (7), 3152-3156 (2013).
18. H. Lourenço-Martins, M. Kociak, S. Meuret, F. Treussart, Y. H. Lee, X. Y. Ling, H.-C. Chang and L. H. Galvão Tizei, ACS Photonics **5** (2), 324-328 (2017).
19. A. Mohtashami and A. Femius Koenderink, New Journal of Physics **15** (4) (2013).

20. R. J. W. Moerland, I. G. Garming, M. W. Kruit, P. Hoogenboom, J. P., Opt Express **24** (21), 24760-24772 (2016).
21. M. A. Schmidt, D. Y. Lei, L. Wondraczek, V. Nazabal and S. A. Maier, Nat Commun **3**, 1108 (2012).
22. M. Sola-Garcia, S. Meuret, T. Coenen and A. Polman, ACS Photonics **7** (1), 232-240 (2020).
23. H. Phillip and E. Taft, Physical Review **136** (5A), A1445 (1964).
24. D. E. Aspnes and A. Studna, Physical review B **27** (2), 985 (1983).
25. G. Liaugaudas, G. Davies, K. Suhling, R. U. A. Khan and D. J. F. Evans, Journal of Physics: Condensed Matter **24** (43) (2012).

# Supporting Information

# Decay rate enhancement of diamond NV-centers on diamond thin films

Hao Li,[*] Jun-Yu Ou, Vassili A. Fedotov and Nikitas Papasimakis

*Optoelectronics Research Center & Center for Photonic Metamaterials, University of Southampton, Southampton SO17 1BJ, UK*

*Correspondence to: H.Li@soton.ac.uk

## S1. Cathodoluminescence experimental setup

Our time-resolved cathodoluminescence system is based on a scanning electron microscope (CamScan CS3000) fitted with a LaB6 electron source, operated at 10 kV in fixed-spot mode in our measurements. The pulsed electron beam was created by a beam blanker driven by a wave function generator (Tektronix AWG7122C). Wavelength and time resolved measurements of emitted photons were performed by a Horiba iHR320 spectrometer and a Horiba TBX picosecond photon detection module, respectively. A Horiba FluoroHub single photon counting controller provided the required synchronization for time-correlated single-photon counting.

## S2.  Fitting of CL measured photon counting histograms

The photon counting histograms (see Fig. 1d in main text) were fitted using the following formula:

$$I = a_1(1 - \delta(t + a_2)) + 0.5 a_1 \delta(t + a_2)\left(e^{-\frac{t+a_2}{\tau_1}} + e^{-\frac{t+a_2}{\tau_2}}\right) + a_3$$

where $a_1$ is the normalized peak photon number, $a_2$ is the sample irradiation time, $a_3$ is the background noise, the faster characteristic time, $\tau_1$, is related to the carrier time[22], while the slower $\tau_2$ corresponds to the $NV^0$ center lifetime.

## S3. ND Cluster size and aspect ratio distributions

The correlation between the recorded lifetimes and ND cluster geometry are examined in Fig. S1 for diamond nanoparticles placed on a diamond film or a bare Si substrate. The decay rates are presented as function of size (Fig. S1(a)) and shape (Fig. S1(b)). In both cases, the decay rate is nearly independent of both ND cluster size and cluster aspect ratio.

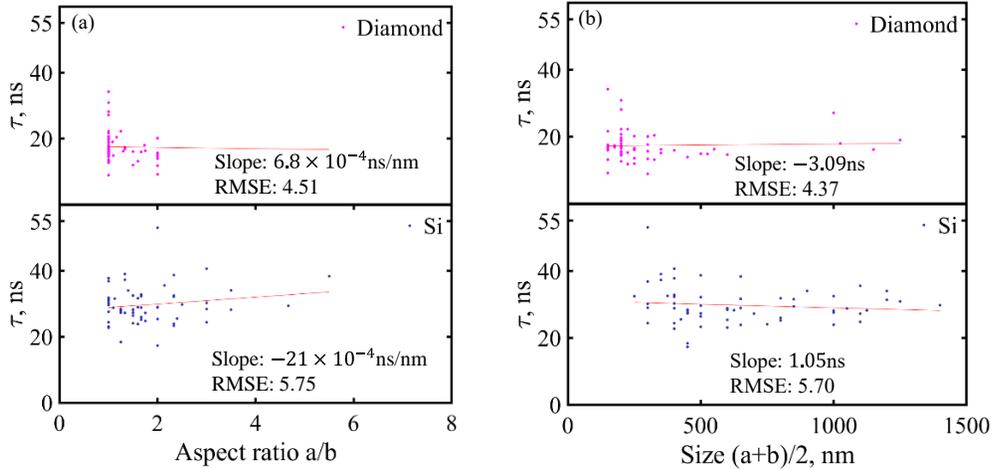

**Fig. S1.** Experimentally measured NV center lifetime vs cluster size (a) and aspect ratio (b) for diamond nanoparticles deposited on a diamond thin film (top) or a silicon substrate (b). The correlation between lifetime and cluster size and aspect ratio is quantified by the slope and RMSE as obtained by a least-squares fit. The cluster size and aspect ratio are calculated as (a+b)/2 and a/b, respectively, where a (b) is the length of the cluster longer (shorter) (see inset to Fig.1(b) in main text).

## S4. Simulation of emission from a nanodiamond with NV center

Numerical simulations were performed by a commercial software (COMSOL 5.3a) based on the finite-element method. We set the refractive index of diamond and Si at the wavelength of 575 nm as $n_{diamond}$=2.40 [23] and $n_{Si}$=4.00+0.03i [24], respectively. The simulation domain represents a cylinder with the radius of 1500 nm and height 3000 nm, terminated by scattering boundaries on all sides (see Fig. S2). The $NV^0$ center excitation at λ=575 nm was introduced via a volume polarization density oscillating inside a 10 nm sphere. The total radiated power, $P_{tot}$, is calculated as the integral of power flow $\vec{P}$ over the surface of a 12 nm large sphere $\vec{S}$ encapsulating the emitting center. The nanodiamond was modeled as a dielectric sphere with radius of 60 nm and with the $NV^0$ center placed at its center. The lifetime of $NV^0$ center on a substrate was given by $\tau = \frac{\tau_{bulk} P_{bulk}}{P_{Tot}}$, where $P_{bulk}$ is the power emitted by the modelled $NV^0$ center in bulk diamond and $\tau_{bulk}$ corresponding to a known bulk lifetime $\tau_{bulk}$=19ns[25]. The total decay rate was calculated as $\gamma_{Tot} = \frac{1}{\tau} = \frac{P_{Tot}}{\tau_{bulk} P_{bulk}}$. The radiative decay rate was obtained by $\gamma_{Rad} = \frac{P_{air}}{\tau_{bulk} P_{bulk}}$, where $P_{air}$ is the power radiated to the far-field. The nonradiative decay rate was calculated as $\gamma_{Non} = \gamma_{Tot} - \gamma_{Rad}$. A summary of the simulated lifetimes and rates for different dipole orientations is shown in Table S1.

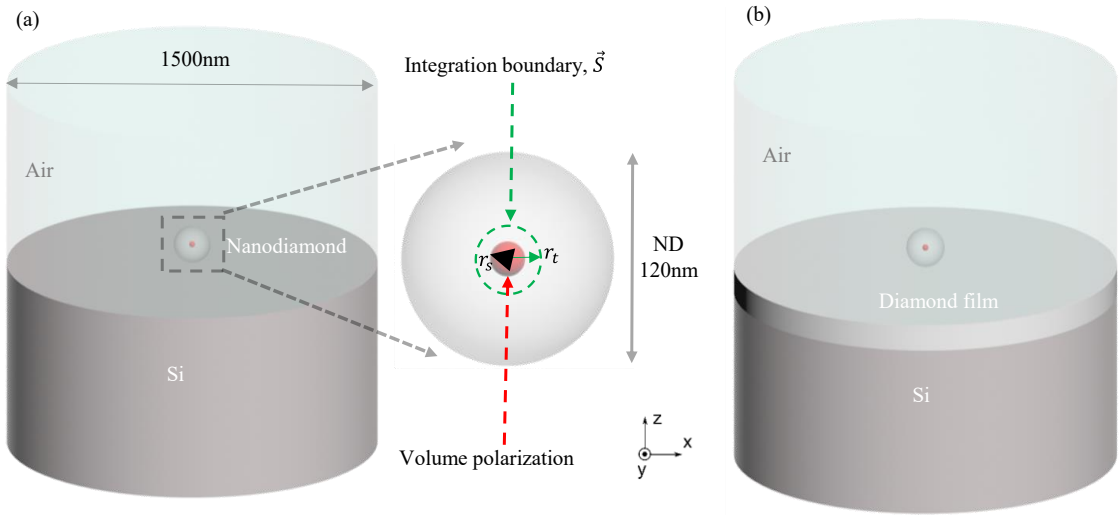

**Fig. S2.** Schematic illustration of the model used in lifetime calculations for a single ND on (a) Si and (b) diamond film on Si. The NV$^0$ center is introduced as a volume polarization density with the orientation along the *x*-axis (in the plane of the substrate) or *z*-axis (normal to the substrate) oscillating inside a sphere of radius $r_s$ =5 nm.

**Table S1. Summary of simulated lifetimes and rates of vertical and horizontal dipole emission and photon loss.**

| Substrates | $\tau_{Sim}^{\parallel}(ns)$ | $\gamma_{Tot}^{\parallel}(\times \mu s^{-1})$ | $\gamma_{Non}^{\parallel}(\times \mu s^{-1})$ | $\gamma_{Rad}^{\parallel}(\times \mu s^{-1})$ |
|---|---|---|---|---|
| Si | 130.55 | 7.66 | 5.30 | 2.36 |
| Diamond | 105.15 | 9.51 | 7.47 | 2.04 |

| Substrates | $\tau_{Sim}^{\perp}(ns)$ | $\gamma_{Tot}^{\perp}(\times \mu s^{-1})$ | $\gamma_{Non}^{\perp}(\times \mu s^{-1})$ | $\gamma_{Rad}^{\perp}(\times \mu s^{-1})$ |
|---|---|---|---|---|
| Si | 38.21 | 26.17 | 17.04 | 9.12 |
| Diamond | 49.31 | 20.28 | 16.33 | 3.95 |

## S5. Collection efficiency of the parabolic mirror

The collection efficiency of the parabolic mirror in our TR-CL system depends on the orientation of the electric dipole of an NV$^0$ center, as shown in Fig. S3. In our experiments, the nanodiamonds are at the focus of the parabolic mirror. We define the quantities $\eta_x^{\parallel} = \frac{P_x}{P_x+P_y+P_z}$, $\eta_y^{\parallel} = \frac{P_y}{P_x+P_y+P_z}$, $\eta_z^{\perp} = \frac{P_z}{P_x+P_y+P_z}$, where $P_x$, $P_y$, $P_z$ are the radiated powers collected by the parabolic mirror for dipoles moments of NV$^0$ centers orientated along x, y, z directions, respectively. The average lifetime can then be calculated as:

$$\bar{\tau}_{Sim} = \frac{1}{\bar{\gamma}_{Tot}} = \frac{1}{(\eta_x^{\parallel} + \eta_y^{\parallel})\gamma_{Tot}^{\parallel} + \eta_z^{\perp}\gamma_{Tot}^{\perp}}$$

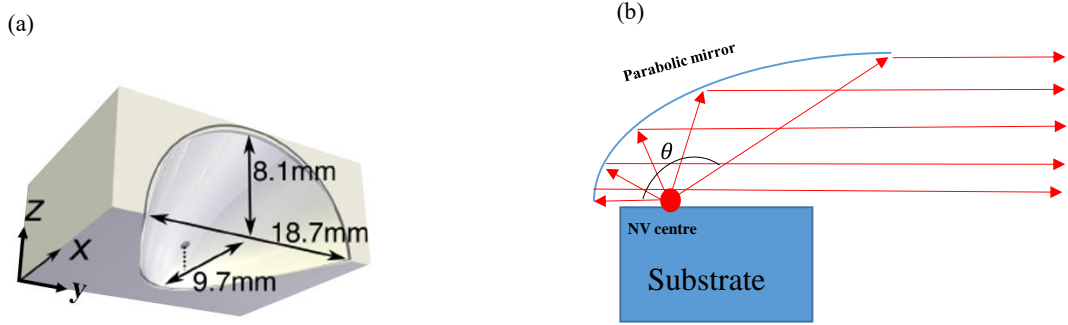

(a)  (b)

**Fig. S3.** (a) Schematic of the parabolic mirror used for collecting photons emitted by NV centers in our experiments. (b) Diagram of photon collection. The angle $\theta$ is ~$135^0$ according to the dimensions in (a).

**Table S2. Collection efficiencies of parabolic mirror for a dipole with x, y z-orientation**

| Substrates | $\eta_x^{\parallel}$ | $\eta_y^{\parallel}$ | $\eta_z^{\perp}$ |
|---|---|---|---|
| Si | 64.04% | 18.29% | 17.67% |
| Diamond | 48.73% | 25.96% | 25.31% |